\documentstyle[epsf]{article}
\def\half{\frac{1}{2}}
\def\be{\begin{equation}}
\def\ba{\begin{eqnarray}}
\def\ea{\end{eqnarray}}
\def\gr{\nabla}
\def\ee{\end{equation}}
\def\to{\rightarrow}
\def\tmu{$TH\epsilon\mu\;$}
\def\pd{\partial}
\def\nn{\not\!}
\def\sp{\overline\psi}
\def\ov{\overline}
\def\api{\frac{\alpha}{\pi}}
\def\al{\alpha}

\def\om{\Omega}
\begin{document}
\onecolumn
\begin{titlepage}
\begin{center}
{\Large \bf The Equivalence Principle and g-2 Experiments} \\ \vspace{2cm}
C. Alvarez\footnotemark\footnotetext{email:
calvarez@avatar.uwaterloo.ca}
and R.B. Mann\footnotemark\footnotetext{email:
mann@avatar.uwaterloo.ca}\\
\vspace{2cm}
Dept. of Physics,
University of Waterloo
Waterloo, ONT N2L 3G1, Canada\\
\vspace{2cm}
\vspace{2cm}
\today\\
\end{center}
\begin{abstract}
We consider the possibility of using measurements
of anomalous magnetic moments of elementary particles as a
possible test of the Einstein Equivalence Principle (EEP).
For the class non-metric theories of gravity  described
by the \tmu formalism we find several novel mechanisms for
breaking the EEP, and discuss the possibilities of setting
new empirical constraints on such effects.

\end{abstract}
\end{titlepage}
\twocolumn

An attractive feature of metric theories of gravity is that they
endow spacetime with a second-rank symmetric tensor field
 $g_{\mu\nu}$ that couples universally to all non-gravitational fields, thereby
affording  a unique
operational spacetime geometry. This feature is a consequence of
the Einstein Equivalence Principle (EEP), which states that the
outcomes of nongravitational test experiments
performed within a local, freely falling frame are independent of
the frame's location (local position invariance, LPI) and velocity
(local Lorentz invariance, LLI)
through a gravitational field.
Non-metric theories of gravity break this universality
by coupling additional gravitational fields to matter, and
so violate either LPI or LLI.
Limits on such effects are imposed by gravitational redshift
and atomic physics experiments
respectively: laser experiments set tight limits on violations of
LLI ($\sim 10^{-22}$) \cite{PLC}, while an upcoming generation of
gravitational redshift experiments could exceed the precision of previous
experiments
\cite{Vessot} by as much as five orders of magnitude \cite{will1}.

In order to empirically check the universal behavior of gravity,
it is important to probe as diverse a range of non-gravitational
interactions as possible for potential
EEP-violating behavior.  To this end,
quantum electrodynamics (QED) provides one such arena in that it allows
one to investigate the behavior of physical systems
whose existence is contingent upon quantum field theoretic
radiative corrections. We began this endeavor
in a previous paper by analyzing the behavior of Lamb shift
transition energies within the
context of nonmetric theories of gravity \cite{catlamb}.
This energy shift, along with anomalous
magnetic moments ($g-2$ ) of fundamental fermions
constitutes the most striking evidence in support of QED
\cite{lamb1,g1}.

We report in this paper the results of an investigation of
the possibility of using measurements
of anomalous magnetic moments of elementary particles as a
possible test of the EEP. Details will appear in a forthcoming paper
\cite{catg2up}.
The high precision attained in $g-2$
experiments motivated earlier work by Newman {\it et al.}
which set new bounds on the validity of special relativity \cite{NFRS}.
We find here that a non-metric spacetime structure induces
qualitatively new effects in the behavior of anomalous magnetic
moments that leave distinctive physical signatures,
allowing the possibility of setting new bounds
on the validity of the EEP.

We follow the approach given in ref.\cite{catlamb}, in which
a gravitationally modified (GM) QED was developed
within  the context of the \tmu formalism \cite{tmu}. This formalism
encompasses a wide class of nonmetric theories of gravity, and deals
with the dynamics of charged particles and electromagnetic fields in
a static, spherically symmetric gravitational field.
The spatial variations of the \tmu functions can be neglected
within atomic scales,
which along with a proper rescaling of coordinates and field
lead to the (GM) QED action:\cite{gabriel}
\be\label{q1}
\!\!S\!=\!\int\!\! d^4x \sp(i\nn\pd+e\nn\! A-m)\psi
 + \half\int\!\! d^4x(E^2-c^2B^2),
\ee
where local natural units are used, $\nn\! A=\gamma_\mu
A^\mu$, $\vec E\equiv-\vec\gr A_0-\pd\vec A/\pd t$,
$\vec B \equiv \vec\gr\times\vec A$
and $c^2=H_0/T_0\epsilon_0\mu_0$ with
the subindex ``0'' denoting the functions evaluated at $\vec X=0$.
The set of parameters $T$, $H$, $\epsilon$, and $\mu$ are
arbitrary functions of the Newtonian gravitational potential
$U= GM/r$, which approaches unity as $U\to 0$, and the metric
is assumed to be
\be\label{tmumet}
ds^2 = T(r)dt^2 - H(r)(dr^2+r^2d\Omega^2) \quad .
\ee

The action (\ref{q1}) has been written with respect
to the preferred frame, as defined by the
rest frame of the external gravitational field $U$. In order to
analyze effects in systems moving with respect to that frame,
we assume that Lorentz transformations relate coordinates
and fields from one system to another, under which the pure
fermion and interaction terms in the action
(\ref{q1}) remain invariant (as do the fermion propagator and the
vertex rule as a consequence).
All nonmetric effects on the moving system
arise from the electromagnetic sector
proportional to $\xi\equiv 1-c^2$ .
This factor is a dimensionless parameter that  scales according to
the magnitude of the dimensionless Newtonian potential,
which turns out to be much smaller than unity for actual experiments.
We are therefore able to compute effects  that break local
Lorentz invariance via a perturbative analysis about the familiar and
well-behaved $c\to 1$ or $\xi\to 0$ limit. Up to $O(\xi)$,
the photon propagator is
(after a proper choice of the gauge fixing term)
\cite{catlamb}:
\begin{equation}\label{11}
G_{\mu\nu}=-(1+\xi)\frac{\eta_{\mu\nu}}{k^2}
+\xi\frac{\gamma^2}{k^2}\left[\eta_{\mu\nu}\frac{(\beta\cdot k)^2}{k^2}
+\beta_\mu\beta_\nu\right] ,
\end{equation}
where $\eta_{\mu\nu}$ is the Minkowski tensor with a signature
(+ - - -);
$\gamma^2\equiv 1/(1-\vec u^2)$, with $\vec u$ as the velocity of the moving
frame with respect to the preferred system, and  $\beta^\mu\equiv(1,\vec u)$;
henceforth $\beta^2\equiv 1-\vec u^2$.

Eq. (\ref{11}) along with the unmodified fermion propagator $S_F(p)$,
and vertex rule form the basis of the Feynman rules of
GMQED. Radiative corrections affecting those quantities are defined
in terms of the photon self energy $\Pi^{\mu\nu}(k)$,
fermion self energy $\Sigma(p)$, and
vertex function $\Gamma^\mu$ respectively.
These insertions involve the calculation
of loop integrals as given by the Feynman rules up to a given order.

We shall consider the lowest order radiative correction
associated with the elastic scattering of
electrons by a static external field $A^\mu$. These one loop
contributions can be summarized in terms of the Feynman diagrams
illustrated in Fig. 1.
\begin{figure}[ht]
\centering
\leavevmode
\epsfxsize=110pt
\epsfbox[90 80 280 360] {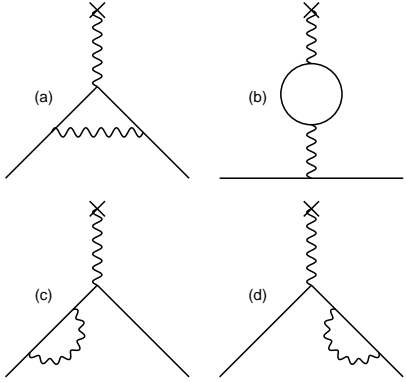}
\caption{One loop corrections for the elastic scattering
of an electron by an external electromagnetic source}
\end{figure}
It is straightforward to check that the
Ward identity $\frac{\pd\Sigma(p)}{\pd p_\mu}=\Gamma^\mu(p,p)$
relating the self energy and vertex function is satisfied.
This result is a consequence of gauge invariance,
and therefore it holds even in the absence of Lorentz invariance.

We have evaluated the amplitudes for the diagrams which
follow from the Feynman rules, giving the total result $\Lambda^\mu$.
To lowest order, the Feynman amplitude related to the elastic
scattering of an electron by a static
external field, is given by $ie\ov u(p')\nn\!\! A(q)u(p)$.
In the nonrelativistic limit of slowly moving particles
($|\vec q|\to 0$) and a static
magnetic field $e\nn A(q)\to -\frac{e}{2m}\vec B\cdot\vec\sigma$.
If we include the radiative corrections ($\Lambda\cdot A$),
we can, up to $O(\xi)O(\al)$,
write the spin magnetic field interaction part as
\be\label{inter}
\!\! H_\sigma\!\!=\!\!-\frac{e}{2m}\!\{\!(g-g_*\vec u^2\!)\!\vec S\cdot\vec B+
g_*\vec S\cdot\vec u\,\vec B\cdot\vec u\}
\!\!\equiv\!\! -\Gamma^{ij}\!S_iB_j
\ee
with
$g\equiv 2+\api[1+\xi\frac{7}{6}]$, and $g_*\equiv-\api\xi\frac{4}{3}$,
where we have identified $\vec S\equiv\frac{\vec\sigma}{2}$,
and $\hat u=\vec u/|\vec u|$.
The presence of preferred frame effects induces a new type of coupling
between the magnetic field and the spin as described by (\ref{inter}).
This interaction stems purely from radiative corrections,
and generalizes the gyromagnetic ratio of a fermion to a tensorial
coupling described by $\Gamma_{ij}$

Hence eq. (\ref{inter}) describes the interaction
(as seen from the particle rest frame) between the particle spin
and an external homogeneous magnetic field. We can extract from this
the energy difference between electrons with opposite
spin projection in the direction of the magnetic field as:
\be\label{eshift}
\Delta E_\sigma=-\frac{eB}{2m}\Big [ g-g_*u^2(1-\cos^2\Theta)\Big]
\ee
where $\Theta$ is the angle between the magnetic field and the
preferred frame velocity. The influence of the radiative corrections
(coming from $g-2$ and $g_*$) in this
energy shift is negligible in comparison to the
dominant factor 2 in $g$.

A more useful means of isolating the effects of these
non-metric corrections
is to study the oscillation of the longitudinal
spin polarization in a magnetic field.
In the metric case, this frequency is proportional to the factor $g-2$,
and so is a signal effect of radiative corrections.

The observable quantity in $g-2$ experiments is actually the electron
polarization, which is proportional to the quantum mechanical
expectation value of $\vec S$. The quantum-mechanical equation of
motion for this value is given by
\be\label{spin1}
\!\!\!\frac{d\vec S}{dt'}\!|_{R.F.}\!\!=\!\!-i[\vec S,\! H_\sigma]\!\!=\!\!\vec
S\!\times\!\frac{e}{2m}\!
\Big[\!(g\!-\!g_*\vec u^2\!)\vec B'\!\!+g_*\!(\vec B'\cdot\vec u)\vec u\Big]
\ee
where the spin expectation value is implicit and the primed variables
are referred explicitly to the particle
rest frame ($R.F.$). Note that preferred frame effects will
distinctly manifest themselves as a temporal variation of the spin
component parallel to the magnetic field.

In general we want to know the spin precession relative
to some specific laboratory system, with respect to which
the particle is moving with some velocity $\vec\beta$.
{\it A-priori} this frame does not need to be the previously defined
preferred frame, and so $\vec\beta\not =\vec u$.

Since the \tmu formalism does not (locally) change the
fermion electromagnetic field interaction, we
assume that a charged particle
in the presence of an homogeneous magnetic field
will still satisfy the equation
$\frac{d\vec\beta}{dt}=\vec\beta\times\vec \om_c$,
with the cyclotron frequency $\vec \om_c=\frac{e}{m\gamma}\vec B$
and $\gamma=(1-\vec\beta^2)^{-1/2}$.
Using Lorentz transformations  to relate the primed
variables in (\ref{spin1}) to the laboratory ones and including the
effect of Thomas precession, we obtain
$d\vec S/dt|_{Lab}=\vec S\times\vec\om_s$
with
$\vec\om_s=\frac{e}{2m}\left[(g-2+2/\gamma-g_*\vec u^2)\vec B+
g_*(\vec B\cdot\vec u)\vec u\right]$,
where we have set $\vec E=0$ and considered the case of orbital
motion perpendicular to the magnetic field ($\vec\beta\cdot\vec B=0$).
Note that the spin precession about $\vec \om_s$ is no longer
parallel to the magnetic field (axial direction), but has a component
parallel to $\vec u$ that comes from radiative and non-metric effects.

At this point it is necessary to define the
preferred coordinate system.
There are several candidates (such as the rest frame of the cosmic
microwave background) for this frame \cite{will1}.
To study this issue it is sufficient
to assume that  the laboratory system (Earth) moves with a
non-relativistic velocity ($\vec V$) with respect to the preferred
frame, and so we can identify
$\vec u=\vec V+\vec\beta$.

In order to single out the effects of radiative corrections, we study
the spin precession relative to the rotational motion of the
electron, that is:
\be\label{set1}
\frac{d\vec S}{dt}|_{rot}=\vec S\times\vec\om_D
\ee
with $\vec\om_D=\vec\om_s-\vec\om_c$.
In the following we refer to the difference
frequency  ($\om_D$) as the anomalous frequency
(given its connection with the anomalous magnetic
moment in the metric case). It is convenient to rewrite
\be\label{set2}
\vec\om_D=\om_a\hat B+\om_a^*\Big[\cos\Theta(\vec V_\bot+\vec\beta)
-2\hat V_\bot\cdot\vec\beta\hat B\Big]
\ee
with
\be\label{anomal}
\om_a=\frac{eB}{2m}\Big[g-2-g_*\Big(\vec V^2+\vec\beta^2
-\vec V^2\cos^2\Theta\Big)\Big]
\ee
and $\om_a^*=\frac{eB}{2m}g_*V$ , where $\Theta$ represents the
angle between  $V$ and the magnetic field, and $V_\bot$ the component
of the velocity perpendicular to $B$. In $\om_a$ we group all the
constant terms parallel to the magnetic field that
contribute to the anomalous frequency (including non-metric effects).
The remaining terms correspond to nonuniform
or off-axial terms (proportional to $\om_a^*$)
that arise from non-metric effects only.

Since $\om_a^*$ is proportional to $\xi$, we can solve for each
component in (\ref{set1})
perturbatively, with $S=S^0+S^*$. Taking, for example,
the initial condition $\vec S(0)=S\hat\beta$
and $\vec V$ on the $XZ$ plane,
we find that the spin components perpendicular to
the magnetic field precess with frequency $\om_a$ about the axial
direction, and that the parallel component follows the motion:
\be
S_\|=S_\|^*=S\frac{\om_a^*}{\om_a}\beta\cos\Theta(\cos \om_at-1)+\cdots
\ee
where we have considered $\om_a<<\om_c$,
and so neglected terms proportional to ($\om_a/\om_c$).

The fact that $\om_a$ was (in the metric case) proportional to
$g-2$ (or $\al$), motivated the very precise $g-2$ experiments which
were designed to measure specifically that anomalous frequency. We see
that this frequency is modified by from its metric value by the
additional terms present in (\ref{anomal}). A comparison
of two electron $g-2$ experiments
(one at electron relativistic energy ($\beta=0.57$) \cite{Ga}
and the other nearly at rest ($\beta=5\times 10^{-5}$) \cite{Gb})
therefore yields the constraint
$|\xi|<10^{-5}$ .
A similar analysis can be carried out for muon $g-2$ experiments, but
the results are less precise than those
for electrons.

Newman {\it et. al.} analyzed these experiments
\cite{NFRS} in order to find
new bounds for the validity of special relativity.
They assumed that the parameter
$\gamma$ involved in the electron motion  had a different value
($\tilde\gamma$) from that which arises kinematically
(in Thomas precession and Lorentz transformations).
The equivalent equation for (\ref{anomal}) is in that case
\be
\Omega_a^{NFRS}=\frac{e
B}{m}\left(\frac{g}{2}-\frac{\gamma}{\tilde\gamma}\right)
\ee
and by comparing with electron $g-2$ experiments, they obtained the
constraint
$\delta \gamma/\tilde\gamma<5.3\times10^{-9} \quad$ .
Our approach is qualitatively
different from theirs, in that we assume $\gamma=\tilde\gamma$
but include preferred frame effects in the evaluation of
the anomalous magnetic moment.

Preferred effects not only modify the anomalous frequency
according to (\ref{anomal}), but also induce oscillations in
the spin component parallel to $B$. As stated above,
this is a qualitatively new signature of EEP violations
due solely to radiative corrections in GMQED. We can estimate this
effect by taking the temporal average
of $S_\|$ over the main oscillation given by $\om_a$, which gives
$\langle S_\|\rangle/S=\frac{2}{3}\xi V\cos\Theta\beta\sim 10^{-25}$,
where we consider a typical experiment with $V\sim 10^{-3}$ and
$\beta\sim 0.5$; and the present constraint for $\xi$.
The novelty of the $S_\|$ oscillation suggests the possibility
of putting tighter constrains on the non-metric parameter,
once appropriate experiments are carried out. The same goes for
the analysis  of $\om_a$ at different values of $\Theta$
(the angle between the magnetic field and the velocity of
the laboratory system with respect to the preferred frame).
The rotation of the Earth will have the effect of converting
this orientation dependence into a time-dependence
of the anomalous magnetic moment, with a period related to that of the
sidereal day.

The previous analysis was concerned with effects related to spatial
anisotropy.   We turn now to a consideration
of possible violations of LPI.
Position dependence in the
former  procedure was implicit in the redefinitions of charge, mass
and fields. These quantities were
rescaled in terms of the local values of the \tmu functions,
which were considered constant throughout the computation.
In LPI violating experiments, a given frequency is measured at two
different points within the same reference system,
where differences in the gravitational potential $\Delta U$
could be significant. The gravitational redshift parameter ($Z$)
accounts for the relative frequency difference via
$Z=\Delta U(1+\Xi)$, where $\Xi$ signal any violation
of LPI related to the frequency involved.
Using standard techniques \cite{will1} we can derive from (\ref{anomal})
the corresponding parameter as
\be
\Xi^a=\frac{11}{6}\Gamma_0-\frac{13}{6}\Lambda_0
\ee
where $ \Gamma_0$ and $\Lambda_0$ parameterize possible LPI violations,
being equal to zero in the metric case. A search
for possible position dependence of anomalous spin precession
frequencies provides another qualitatively new test of
LPI that is sensitive to radiative corrections.

Refined measurements of anomalous magnetic moments can provide
an interesting new arena for investigating the validity of the EEP
in physical systems where radiative corrections are important.
We have considered this possibility explicitly for the class of
non-metric theories described by the \tmu formalism.
The non-universal character of the gravitational couplings in such
theories  affects
the one loop corrections to the scattering amplitude of a free
fermion in an external electromagnetic field
in a rather complicated way, giving rise to several novel effects.
It will be a challenge to set new empirical bounds on such effects
in the next generation of experiments.

We are grateful to M. Haugan for drawing our attention to ref.
\cite{NFRS}.
This work was supported in part by the Natural Sciences and Engineering
Research Council of Canada.

\end{document}